\documentclass[aps,groupedaddress, twocolumn,  nofootinbib]{revtex4-1}
\usepackage{latexsym}
\usepackage{amsmath}
\usepackage{amsfonts}
\usepackage{amsbsy}
\usepackage{amssymb}
\usepackage{epsfig}
\usepackage{mathrsfs}
\usepackage{epsfig}
\usepackage{bbm}
\usepackage{color}

\begin{document}

\title{{\bf Deformation of the EPRL spin foam model by a cosmological constant}}

\author{Benjamin Bahr}
\email[]{benjamin.bahr@desy.de}
\author{Giovanni Rabuffo}
\email[]{giovanni.rabuffo@desy.de}
\affiliation{II. Institute for Theoretical Physics\\University of Hamburg\\ Luruper Chaussee 149\\22761 Hamburg\\Germany}


\date{\today}

\begin{abstract}
In this article, we consider an ad-hoc deformation of the EPRL model for quantum gravity by a cosmological constant term. This sort of deformation has been first introduced by Han for the case of the $4$-simplex. In this article, we generalise the deformation to the case of arbitrary vertices, and compute its large-$j$-asymptotics. We show that, if the boundary data corresponds to a $4d$ polyhedron $P$, then the asymptotic formula gives the usual Regge action plus a cosmological constant term. We pay particular attention to the determinant of the Hessian matrix, and show that it can be related to the one of the  undeformed vertex. 

\end{abstract}


\maketitle
\section{Motivation}

Spin Foam models are tentative proposals for a path integral formulation of quantum gravity. They are a very active research subject, and have many connection points with state sum models, tensor field theories, and loop quantum gravity (\cite{Perez:2012wv} and references therein). 

One of the most widely studied model ist the one by Engle, Pereira, Rovelli and Livine \cite{Engle:2007wy}.\footnote{Other models include the one by Barrett and Crane \cite{Barrett:1997gw}, Freidel and Krasnov \cite{Freidel:2007py}, Baratin and Thiemann \cite{CubulationSpinFoamThiemann2008}, Baratin and Oriti \cite{Baratin:2011hp}. } It provides the definition of a so-called vertex amplitude $\mathcal{A}_v$, which assigns a transition amplitude to a spin network state, which is interpreted as 3d boundary geometry. The boundary is that of a small piece of $4d$ space-time (a ``vertex''), while the whole path integral is defined by glueing many of these vertices to one large $2$-complex (e.g.~described in \cite{Bahr:2010bs, Kisielowski:2011vu, Bahr:2012qj}). 

Each of the local boundary spin network states is defined on a graph $\Gamma$. The spin network states on $\Gamma$ form a Hilbert space $\mathcal{H}_{\Gamma}$, and the amplitude can be regarded as linear form on this space. The original EPRL amplitude was defined on a complete graph $K_5$, corresponding to the boundary of a $4$-simplex. The model has been generalized to arbitrary graphs \cite{Kaminski:2009fm}, although it can be argued that the model would have to be amended to include the correct implementation of the volume simplicity constraints \cite{Belov:2017who, Bahr:2017ajs}. 

There is an asymptotic regime of the amplitude, in which one can show it to be connected to the exponential of the Regge action, i.e.~a discrete analogue of the Einstein Hilbert action for general relativity \cite{Barrett:2009gg}. This, among others, is an indication of the connection between the EPRL model and the path integral for quantum gravity.

There are several versions of deformation of this model to include a non-zero cosmological constant. The most technically clean one is a deformation of the underlying group $SU(2)$ to a quantum group $SU(2)_q$, with $q=e^{ \frac{2\pi i}{k+2}}$ a root of unity, where $\Lambda =6\pi/(\ell_P^2 k)$. \cite{Fairbairn:2010cp, Han:2011aa, Haggard:2015kew, Dittrich:2018dvs}. One of the earliest deformations of the model, however, was still on the level of classical groups, by deforming $\mathcal{A}_\Gamma$, keeping $\mathcal{H}_\Gamma$ unchanged. The definition was given by Han \cite{Han:2011aa}, for the case of a $4$-simplex, and a partial analysis of the asymptotic regime was given, which demonstrated the emergence of the Regge action plus a cosmological constant term. 

While this ad-hoc deformation of the EPRL model shows no obvious connection to the later definitions with quantum groups, it is a useful tool for calculations. In particular, in recent calculations concerning the RG flow of the EPRL model (see \cite{Bahr:2016hwc, Bahr:2017eyi, Bahr:2017klw}), it turned out to be desirable to have a running cosmological constant. The boundary graphs are more complicated in that case, so a generalisation of Han's deformation to more complicated graphs is needed. This is what is going to be undertaken in this article.


The plan of this article is as follows: First, we will remind the reader of the definition of the undeformed EPRL amplitude in section \ref{Sec:Undeformed}. We will then give a general definition of the deformed amplitude in section \ref{Sec:Deformation}. This will be a straightforward generalisation of Han's formula. The main part of the article will be in the following section \ref{Sec:Asymptotics}, where we will consider the asymptotic analysis of this amplitude in the general case. In particular, we will prove the (highly non-trivial) statement, that the asymptotic of the deformed amplitude coincides with the undeformed one, apart from a cosmological constant term. This requires a very careful handling of the determinant of the Hessian matrix in the stationary phase approximation, and we will divert some of the (rather technical) details to the appendix \ref{App:MatricesAndDeterminants}.

\newpage
\section{The undeformed model}\label{Sec:Undeformed}

\begin{figure}[hbt!]
\includegraphics[scale=0.9]{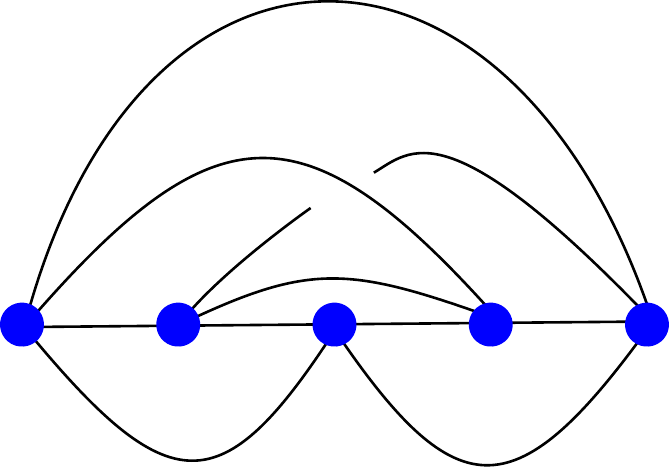}
\caption{The boundary graph of a 4-simplex.}\label{Fig:4SimplexGraph}
\end{figure}

We consider a general spin foam vertex, for the Riemannian signature EPRL-FK model, for Barbero-Immirzi paramter $\gamma\in(0, 1)$. The amplitude is a linear map on the boundary Hilbert space. A state in that Hilbert space is given by boundary data, which is completely described by a directed graph $\Gamma\subset S^3$ embedded into a three-sphere. For example, for a 4-simplex the graph is given by the complete graph with five nodes, with the knotting as in figure \ref{Fig:4SimplexGraph}. \footnote{In general, it is suspected that the knotting of the graph is important for the formula of the vertex amplitude, as soon as quantum groups are involved. As it turns out, however, Han's heuristic deformation does not seem to depend on the precise knotting class of the graph $\Gamma$.}

A boundary geometry on $\Gamma$ is given by a collection of spins $j_L\in\frac{1}{2}\mathbb{N}$ associated to the links $L\in \text{Links}(\Gamma)$ of $\Gamma$, and a collection of $3d$ unit vectors $n_{NL}$ associated to pairs of nodes $N\in \text{Nodes}(\Gamma)$ of the graph, and links $L$ which are connected to $N$. For all $L\supset N$, the corresponding unit vectors are chosen such that they satisfy
\begin{eqnarray}\label{Eq:Closure}
G_N\;:=\;\sum_{L\leftarrow N}j_L {n}_{NL}\;-\;\sum_{L\rightarrow N}j_L {n}_{NL}\;=\;0.
\end{eqnarray}

\noindent Here, the notation $L\rightarrow N$ denotes all links which are ingoing to the node $N$, while $L\leftarrow N$ denotes the outgoing links.

For Riemannian signature, the local gauge group is $\text{Spin}(4)$. We use the Hodge duality in four dimensions, under which its Lie algebra decomposes into $\mathfrak{spin}(4)\simeq \mathfrak{su}(2)\oplus\mathfrak{su}(2)$, two commuting $SU(2)$-subalgebras, which are the eigenspaces under the Hodge $*$ for eigenvalues $\pm 1$. Consequently, one has the group isomorphism $\text{Spin}(4)\simeq SU(2)\times SU(2)$, and an irreducible representation of $\text{Spin}(4)$ can therefore be depicted as pair $(j^+,j^-)$ of half-integers.

The undeformed vertex amplitude $\mathcal{A}_\Gamma$ is constructed in the following way: Define\footnote{With the definition (\ref{Eq:SpinsSimplicity}), one has to demand that all three $j_L, j_L^\pm$ are half-integers, which puts severe restrictions on the Barbero-Immirzi parameter $\gamma$. This is a pathology of the Riemannian model, which does not occur in the Lorentzian context.}
\begin{eqnarray}\label{Eq:SpinsSimplicity}
j^\pm_L\;:=\;\frac{|1\pm \gamma|}{2}j_L.
\end{eqnarray}

\noindent For a unit vector  ${n}\in S^2$, define the coherent states
\begin{eqnarray}
|j,\,{n}\rangle\;:=\;D_j(g_{{n}})|j\;j\rangle,
\end{eqnarray}

\noindent i.e.~the action on the highest weight vector with a group element $g_{{n}}$, which is such that $g_{{n}}e_z={n}$, with $e_z$ being the unit vector in $z$-direction.\footnote{Given ${n}\in S^2$, the corresponding $g_{{n}}$ is only defined up to a $U(1)\subset SU(2)$-subgroup. Different choices amount to states $|j,\,{n}\rangle$ which differ by a complex phase. For one vertex amplitude, this phase is not important, while for larger triangulations, the relative phases of these states in neighbouring vertices have to be taken care of, since they encode the $4d$ curvature. }

Define the boosting map
\begin{eqnarray}
\beta_L : V_{j_L}\;\longrightarrow\;V_{j_L^+}\otimes V_{j_L^-}
\end{eqnarray}

\noindent is the isometric embedding of $j_L$ into the corresponding subspace of the Clebsh-Gordon decomposition of
\begin{eqnarray}\label{Eq:LinkHilbertSpaces}
V_{j_L^+,j_L^-}\simeq V_{j_L^+}\otimes V_{j_L^-}\simeq V_{|j_L^+-j_L^-|}\oplus \cdots V_{j_L^++j_L^-}.
\end{eqnarray}

\noindent Also, denote by $\mathcal{P}:\mathcal{H}\to\text{Inv}_{SU(2)\times SU(2)}(\mathcal{H})$ the operator
\begin{eqnarray}
\mathcal{H}\;:=\;\left(\bigotimes_{L\leftarrow N}V_{j_L^+}\otimes V_{j_L^-}\right)\otimes \left(\bigotimes_{L\rightarrow N}V_{j_L^+}^\dag\otimes V_{j_L^-}^\dag\right),
\end{eqnarray}

\noindent which is the projector onto the invariant subspace of the Hilbert space $\mathcal{H}$.

\noindent With this, one defines the boosted \emph{Livine-Speziale-intertwiners}
\begin{eqnarray}\label{Eq:BoostedIntertwiner}
\iota_N^\pm\;:=\;\mathcal{P}\left[\bigotimes_{L\leftarrow N} \beta_L|j_L,\,{n}_{NL}\rangle
\otimes\bigotimes_{L\rightarrow N}\langle j_L,\,{n}_{NL}|\beta_L^\dag \right].
\end{eqnarray}

\noindent As a result of this definition, the tensor product of all boosted Livine-Speziale intertwiners (\ref{Eq:BoostedIntertwiner}) is an endomorphism on the tensor product of all representation spaces over the links, i.e.~
\begin{eqnarray}
\bigotimes_N\;\iota_N^\pm\;:\;\bigotimes_L\left(V_{j_L^+}\otimes V_{j_L^-}\right)\;\longrightarrow\;\bigotimes_L\left(V_{j_L^+}\otimes V_{j_L^-}\right),
\end{eqnarray}

\noindent where the $\iota^\pm=(\iota^+,\iota^-)$ factorise due to $\gamma<1$. The vertex amplitude $\mathcal{A}_v$ is defined as the trace of this map, i.e.~
\begin{eqnarray}\label{Eq:VertexTrace}
\mathcal{A}_v\;:=\;\text{tr}\left(\bigotimes_N\;\iota_N^\pm\right)\;=\;\mathcal{A}^+_v\mathcal{A}^-_v.
\end{eqnarray}

\section{Deformation of the model}\label{Sec:Deformation}

We now deform the model with a cosmological constant term.  The state-of-the-art method to do this is to resort to replacing the group $SU(2)$, which features prominently in the construction of the EPRL-FK model, by its quantum group counterpart $SU(2)_q$, with $q=e^{ \frac{2\pi i}{k+2}}$ a root of unity, where $\Lambda =6\pi/(\ell_P^2 k)$.

There is a heuristic alternative to this, which relies on a deformation of the EPRL-FK model, which stays purely on the classical level. This was proposed by Han in the case of a 4-simplex \cite{Han:2011aa}. Here we generalise Han's result to arbitrary vertices, and perform the large-$j$-asymptotics, including the treatment of the Hessian matrix.

\begin{figure}
\includegraphics[scale=0.9]{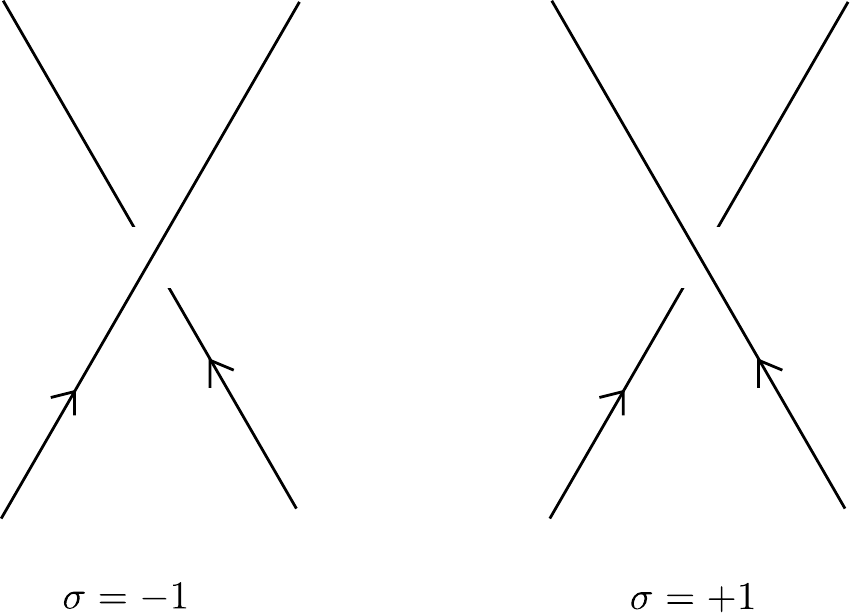}
\caption{The two types of crossings $C$ get assigned different numbers $\sigma(C)=\pm 1$.}\label{Fig:TypesOfCrossings}
\end{figure}

Given the definition of the vertex amplitude $\mathcal{A}_\Gamma$, the deformation is given in terms of a parameter $\omega\in\mathbb{R}$. It is constructed as follows: The graph $\Gamma$ needs to be projected down to the $2d$ plane, where it can be depicted with \emph{crossings} (see figure \ref{Fig:HypercubicGraph} for the example of a hypercubic graph). For each crossing $C$ in the graph between two links $L, L'$ with spins $k_L$, $k_{L'}$, define the crossing operator
\begin{eqnarray}
R_C\;:=\;e^{i \omega \sigma(C) V_C}
\end{eqnarray}

\noindent were $\omega\in\mathbb{R}$ is the deformation parameter,  $\sigma(C)=\pm 1$ is the type of crossing (ove- or under-crossing, see figure \ref{Fig:TypesOfCrossings}), and with

\begin{eqnarray}
V_C\;:=\;\sum_{\epsilon = \pm}\frac{\epsilon 4 }{(1\,\epsilon\, \gamma)^2}\sum_{I=1}^3 D_{(j_L^\epsilon)}(X_I^\epsilon)\otimes D_{(j_{L'}^\epsilon)}(X_I^\epsilon),
\end{eqnarray}

\noindent where the $X_I^+$ ($X_I^-$) are an orthonormal basis of the self-dual (anti-self-dual) $\mathfrak{su}(2)$. The operator $R_C$ acts as endomorphism on $V_{(j_L^+,j_L^-)}\otimes V_{(j_{L'}^+,j_{L'}^-)}$ (\ref{Eq:LinkHilbertSpaces}). By tensoring $\otimes_C R_C$ with the identity operator for all links in $\Gamma$ which do not appear in a crossing, we obtain an endomorphism on $\mathcal{H}$\footnote{Technically, by this definition, the graph has to be such that each link in the graph $\Gamma$ is part of at most one crossing. This is of course not the most general case, but our definition can be straightforwardly extended to a graph with arbitrarily many crossings per link, by trivially subdividing that link with two-valent vertices, onto which one places the unique (by Schur's lemma) normalized intertwiner.}. The deformed vertex amplitude $\mathcal{A}^\omega_\Gamma$ is then defined as
\begin{eqnarray}
\mathcal{A}_\Gamma^\omega\;:=\;\text{tr}\left(\bigotimes_N\iota_N^\pm\;\Big(\mathbbm{1}\otimes \bigotimes_C R_C\Big)\right).
\end{eqnarray}

\noindent Note that while $R_C$ depends on the choice of orthonormal basis, the amplitude $\mathcal{A}_\Gamma^\omega$ does not, due to the gauge-invariance of each boosted Livine-Speziale intertwiner.

\section{Large-$j$ asymptotics of the deformed amplitude}\label{Sec:Asymptotics}

In the case of $\gamma\in(0, 1)$, the undeformed amplitude $\mathcal{A}_\Gamma=\mathcal{A}_\Gamma^+\mathcal{A}_\Gamma^-$ factorises over the two sectors (selfdual and anti-selfdual). Since the respective generators $[X_I^+,\,X_J^-]=0$ commute, so do the $R_C=R_C^+R_C^-$, of course. Hence, also the deformed amplitude factorises:
\begin{eqnarray}
\mathcal{A}_\Gamma^\omega\;=\mathcal{A}_\Gamma^{\omega,+}\mathcal{A}_\Gamma^{\omega,-}.
\end{eqnarray}

\noindent First we note that, due to the factorisation property, it is enough to look at only the $+$-part. To simplify notation, in what follows we abbreviate $j_L^+\to j$, $j_{L'}^+\to j'$, $D_{J^+_L}(X_I^\pm)\to X_I$, $g_a^+\to g_a$, etc..

In particular, we have that the (undeformed) $+$-amplitude is given by

\begin{eqnarray}\nonumber
\mathcal{A}_\Gamma^+\;=\;\int_{SU(2)^{N_\Gamma}}dg_a\; \prod_{b\to a}\; \langle j_{ab},\,n_{ab}|\;(g_a)^{-1}\,g_b\;|j_{ab},\,n_{ba}\rangle\\[5pt] \label{Eq:VertexAmplitudeFactorised}
\end{eqnarray}

\noindent where the product ranges over all links, where in the formula $b$ is the starting point (source) of the link, and $a$ is the end point (target).

Now assume that there is a crossing between the link $b\to a$ and $b'\to a'$. Then, in the deformed amplitude, the two corresponding factors in the product (\ref{Eq:VertexAmplitudeFactorised}) are replaced by
\begin{eqnarray}\label{Eq:DeformedPropagators}
\langle\Psi|\;\exp\left(\frac{4i\omega\sigma(C)}{(1+\gamma)^2}\sum_{I=1}^3X_I\otimes X_I \right)\;|\Phi\rangle
\end{eqnarray}

\noindent with
\begin{eqnarray*}
\langle\Psi|\;&=&\;\langle j_{ab},\,n_{ab}|(g_a)^{-1}\otimes \langle j_{a'b'},\,n_{a'b'}|(g_{a'})^{-1}\\[5pt]
|\Phi\rangle\;&=&\;g_b\;|j_{ab},\,n_{ba}\rangle\otimes g_{b'}\;|j_{a'b'},\,n_{b'a'}\rangle
\end{eqnarray*}
\noindent The expression (\ref{Eq:DeformedPropagators}) can be expanded to
\begin{widetext}
\begin{eqnarray}\label{Eq:ExpansionOfExponential}
\sum_{n=0}^\infty\frac{1}{n!}\left(\frac{4i\omega}{(1+\gamma)^2}\right)^n\sum_{I_1,I_2,\ldots, I_n=1}^3
&\langle j_{ab},\,n_{ab}|\;(g_a)^{-1}\,X_{I_1}X_{I_2}\cdots X_{I_n}\,g_b\;|j_{ab},\,n_{ba}\rangle\\[5pt]\nonumber
\times &\langle j_{a'b'},\,n_{a'b'}|\;(g_{a'})^{-1}\,X_{I_1}X_{I_2}\cdots X_{I_n}\,g_{b'}\;|j_{a'b'},\,n_{b'a'}\rangle
\end{eqnarray}
\end{widetext}


\noindent To consider the stationary phase of an individual term, we use the resolution of identity
\begin{eqnarray}
(2j+1)\int_{S^2}d^2n\;|j,\,n\rangle\langle j,\,n|\;=\;\mathbbm{1}_{V_j}
\end{eqnarray}

\noindent $n-1$ times, and write
\begin{eqnarray*}
&&\hspace{-25pt}\langle j_{ab},\,n_{ab}|\;(g_a)^{-1}\,X_{I_1}X_{I_2}\cdots X_{I_n}\,g_b\;|j_{ab},\,n_{ba}\rangle\\[5pt]
\;&=&\;(2j+1)^{n-1}\int_{(S^2)^{n-1}}d^2n_i\;\langle j,\,n_{ab}|(g_a)^{-1}X_{I_1}|j,\,n_1\rangle\\[5pt]
&&\times
\langle j,\,n_1|X_{I_2}|j,\,n_2\rangle\cdots\langle j,\,n_{n-1}|X_{I_{n-1}}g_b|j, \,n_{ba}\rangle
\end{eqnarray*}

\noindent The $X_I$ are the generators of the $\mathfrak{su}(2)$ Lie algebra $[X_I,X_J]=i\epsilon_{IJK}X_K$, which is why, in the spin-$\frac{1}{2}$-representation, we have $X_I\;=\;\sigma_I/2$ in terms of the Pauli matrices $\sigma_I$. We have therefore
\begin{eqnarray}\label{Eq:SomeEquation}
\langle j,\,n|X_I|j,\,n'\rangle\;=\;j\langle n|\sigma_I|n'\rangle\;\langle n|n'\rangle^{2j-1},
\end{eqnarray}

\noindent where $|n\rangle:=|\frac{1}{2},\,n\rangle$. \footnote{One can show (\ref{Eq:SomeEquation}) easily by using $X_I=-i\frac{d}{dt}_{|t=0}e^{i t X_I}$ and the product rule.} With this, we can write

\begin{eqnarray}\label{Eq:SOMETERM}
&&\langle j_{ab},\,n_{ab}|\;(g_a)^{-1}\,X_{I_1}X_{I_2}\cdots X_{I_n}\,g_b\;|j_{ab},\,n_{ba}\rangle\\[5pt]
&&\;=\int_{(S^2)^{n-1}}d^2n_i\;a(n_i, g_a, g_b)\;e^{S(n_i, g_a, g_b)}
\end{eqnarray}

\noindent with

\begin{widetext}
\begin{eqnarray*}
a(n_i, g_a, g_b)\;&=&\;(2j+1)^{n-1}j^n\frac{\langle n_{ab}|(g_a)^{-1}\sigma_{I_1}|n_1\rangle}{\langle n_{ab}|(g_a)^{-1}|n_1\rangle}
\frac{\langle n_1|\sigma_{I_2}|n_2\rangle}{\langle n_1|n_2\rangle}\cdots
\frac{\langle n_{n-1}|\sigma_{I_{n}}g_b|n_{ba}\rangle}{\langle n_{{n-1}}|g_b|n_{ba}\rangle}
\\[5pt]
S(n_i, g_a, g_b)\;&=&\;2j\Big(\ln \langle n_{ab}|(g_a)^{-1}|n_1\rangle+\ln\langle n_1|n_2\rangle+\cdots +\ln\langle n_{n-1}|g_b|n_{ba}\rangle\Big).
\end{eqnarray*}
\end{widetext}

\noindent This is now in a form where one can perform the (extended) stationary phase approximation. Note that this is for one term in the sum (\ref{Eq:ExpansionOfExponential}) only, and the variables are all the $g_a$, and, for every crossing, $n_i$ and $n'_i$ with $i=1,\ldots, n-1$ (the $n'_i$ vectors come from the term similar to (\ref{Eq:SOMETERM}), with the dashed nodes $a'$, $b'$). First, we note that the criticality condition $\text{Re}S=0$ (where we consider the whole action for $\mathcal{A}_\Gamma^\omega$ now), is equivalent to
\begin{eqnarray}\label{Eq:CritStat01}
g_an_{ab}\;=\;g_bn_{ba}
\end{eqnarray}

\noindent and
\begin{eqnarray}\label{Eq:CritStat02}
n_i=g_bn_{ba},\qquad n'_i=g_{b'}n_{b'a'}\text{ for all }i.
\end{eqnarray}

\noindent One should note that the criticality equations (\ref{Eq:CritStat01}) for the group elements $g_a$ are precisely the ones for the \emph{undeformed} amplitude. The criticality equations for the unit vectors $n_i$, $n'_i$ (remember that, per crossing, there are $2(n-1)$ unit vectors), are such that, on each edge which participates in some crossing, all vectors have to be equal, and coincide with the two normal vectors $g_an_{ab}=g_bn_{ba}$. This shows that, using the same gauge symmetry as in the undeformed case, setting one $g_a=\mathbbm{1}$, all critical points are isolated, when they are also isolated in the undeformed case.\footnote{This is, indeed, the generic case, e.g~in the case of the $n_{ab}$ forming a Regge boundary geometry at the $4$-simplex \cite{Barrett:2009gg}, in all cases of the hypercuboid \cite{Bahr:2015gxa}, or the hyperfrustum, if $\alpha\in(\pi/4, 3\pi/4)$. \cite{Bahr:2017eyi}}

The stationary points are equally easily identified, and they are, as in the undeformed case, the closure condition for each node, and $n_i=g_bn_{ba}$, $n'_i=g_{b'}n_{b'a'}$ for all $i$.

In particular, this means that, after gauge-fixing, the critical and stationary points of the deformed and the undeformed amplitude are in one-to-one correspondence. Furthermore, it is easy to see that the value of the respective actions, evaluated at corresponding critical stationary points, coincide.

\subsection{The Hessian matrix}

To make the notation easier, we assume that there is only one crossing. The general case with many crossings can be treated similarly, though. We also assume that there is at least one critical, stationary point $g_a^{(c)}$ for the undeformed (gauge-fixed) amplitude. Before we continue, we perform a coordinate transformation on the $g_a,n_i,n'_i$ variables, via
\begin{eqnarray}
g_a\to g_a,\qquad n_i\to g_b n_i,\qquad n'_i\to g_{b'}n'_i.
\end{eqnarray}

\noindent Since $SU(2)$ acts via rotations on $S^2$, the Jacobi matrix for this transformations is equal to unity. The action after the coordinate transformation is then given by

\begin{widetext}
\begin{eqnarray}\nonumber
S(g_a,n_i,n'_i)\;&=&\;\sum_{cd\neq ab, a'b'}2j_{cd}\ln \langle n_{cd}|(g_c)^{-1}g_d|n_{dc}\rangle\\[5pt]\label{Eq:FinalAction}
&+&2j_{ab}\Big(\ln \langle n_{ab}|(g_a)^{-1}g_b|n_1\rangle+\ln\langle n_1|n_2\rangle+\cdots +\ln\langle n_{n-1}|n_{ba}\rangle\Big)\\[5pt]\nonumber
&+&2j_{a'b'}\Big(\ln \langle n_{a'b'}|(g_{a'})^{-1}g_{b'}|n'_1\rangle+\ln\langle n'_1|n'_2\rangle+\cdots +\ln\langle n'_{n-1}|n_{b'a'}\rangle\Big).
\end{eqnarray}
\end{widetext}

Note that the first two lines in (\ref{Eq:FinalAction}) are the same as in the undeformed case, while the remaining two come from the deformation due to the crossing. We compute the Hessian matrix for the deformed amplitude, at the critical stationary point
\begin{eqnarray}
g_a\;=\;g_a^{ (c)},\qquad n_{i} = n_{ba},\qquad n'_{i} = n_{b'a'}.
\end{eqnarray}

\noindent In particular, we introduce coordinates around this point via $g_a\;=\;e^{i x_a^I\sigma_I}g_a^{(c)}$ and
\begin{eqnarray}\nonumber
n_i=g_{n_{ba}}
\left(
\begin{array}{c}
\sin \theta_i\\
\sin\phi_i\,\cos\theta_i\\
\cos\phi_i\,\cos\theta_i
\end{array}
\right)
,
 n'_i=g_{n_{b'a'}}
\left(
\begin{array}{c}
\sin \chi_i\\
\sin\xi_i\,\cos\chi_i\\
\cos\xi_i\,\cos\chi_i
\end{array}
\right),\\[5pt]\label{Eq:ParameterizationOfSphere}
\end{eqnarray}

\noindent where the angles take values in $\phi_i,\xi_i\in(-\pi,\pi)$ and $\theta_i,\chi_i\in(-\frac{\pi}{2},\frac{\pi}{2})$. The critical and stationary point is assumed at $x_I^a=0,\phi_i=\xi_i=\theta_i=\chi_i=0$. The vectors $|n_i\rangle$ are then given by
\begin{eqnarray*}
|n_i\rangle\;&=&\;g_{n_{ba}}\,\exp\left(i\frac{\phi_i}{2}\sigma_1\right)\exp\left(-i\frac{\theta_i}{2}\sigma_2\right)|e_z\rangle\\[5pt]
&=&\;g_{n_{ba}}\,\left[\left(\cos\frac{\phi_i}{2}\cos\frac{\theta_i}{2}+i\sin \frac{\phi_i}{2}\sin\frac{\theta_i}{2}\right)|\uparrow\rangle\right.\\[5pt]
&&\qquad\left.+\left(\cos\frac{\phi_i}{2}\sin\frac{\theta_i}{2}+i\sin \frac{\phi_i}{2}\cos\frac{\theta_i}{2}\right)|\downarrow\rangle\right],
\end{eqnarray*}

\noindent where $|e_z\rangle=|\uparrow\rangle$ is the highest weight vector in the spin $\frac{1}{2}$-representation. A similar formula holds for $|n_i'\rangle$.  This leads to
\begin{eqnarray}
\frac{\partial}{\partial \phi_i}\langle n_i|n_{i+1}\rangle_{|\text{crit, stat}}\;=\;0,
\end{eqnarray}

\noindent and similar relations for the other angles. Also, 

\begin{eqnarray}
\langle n_i|n_{i+1}\rangle_{|\text{crit, stat}} = 1.
\end{eqnarray}

Therefore, for all second derivatives which have at least one derivative w.r.t.~one of the angles, the $\ln$ can be left out, e.g.:
\begin{eqnarray*}
&&\frac{\partial^2}{\partial \phi_i\partial\phi_{i+1}}\ln\langle n_i|n_{i+1}\rangle_{|\text{crit, stat}}\\[5pt]
&&\;=\;
\frac{\partial^2}{\partial \phi_i\partial\phi_{i+1}}\langle n_i|n_{i+1}\rangle_{|\text{crit, stat}},
\end{eqnarray*}

\noindent and similar relations for all other varying types of angles. Thus we get, at the stationary and critical points:
\begin{eqnarray}\label{Eq:MatrixElement_01}
\frac{\partial^2S}{\partial \phi_i^2}\;&=&\;
\frac{\partial^2S}{\partial \theta_{i}^2}\;=\;-j_{ab},\\[5pt]\nonumber
\frac{\partial^2S}{\partial \xi_i^2}\;&=&\;
\frac{\partial^2S}{\partial \chi_{i}^2}\;=\;-j_{a'b'}.
\end{eqnarray}

\noindent Also, we get
\begin{eqnarray}\nonumber
\frac{\partial^2S}{\partial \theta_i\partial\theta_{i+1}}\;&=&\;
\frac{\partial^2S}{\partial \phi_i\partial\phi_{i+1}}\;=\;\frac{j_{ab}}{2},\qquad
\frac{\partial^2S}{\partial \phi_i\partial\theta_{i}}\;=\;0,\\[5pt]\label{Eq:MatrixElement_02}
\frac{\partial^2S}{\partial \phi_i\partial\theta_{i+1}}\;&=&\;i\frac{j_{ab}}{2},\qquad
\frac{\partial^2S}{\partial \phi_{i+1}\partial\theta_{i}}\;=\;-i\frac{\,j_{ab}}{2}.
\end{eqnarray}

\noindent All other mixed $\phi, \theta$ angle derivatives are zero. For $\xi,\chi$ angles similar relations hold. Furthermore, we have
\begin{eqnarray}
\langle n_{ab}|(g_a)^{-1}g_b|n_{1}\rangle_{|\text{crit, stat}}\;=\;e^{i\psi}.
\end{eqnarray}

\noindent Using this and $g_ag_{n_{ab}}=g_bg_{n_{ba}}e^{-i\psi \sigma_3}$, we get on the critical and stationary point that

\begin{widetext}
\begin{eqnarray}\nonumber
\frac{\partial^2S}{\partial x_b^I\partial \phi_1}\;&=&\;2j_{ab}\frac{\partial^2}{\partial x_b^I\partial \phi_1}\ln\langle n_{ab}|(g_a)^{-1}e^{i x_b^J\sigma_J}g_bg_{n_{ba}}e^{i\phi_1\sigma_1/2}e^{-i\theta_1\sigma_2/2}|\uparrow\rangle\\[5pt]\label{Eq:MatrixElement_03}
&=&\;2j_{ab}e^{-i\psi}\frac{\partial^2}{\partial x_b^I\partial \phi_1}\langle n_{ab}|(g_a)^{-1}e^{i x_b^I\sigma_I}g_bg_{n_{ba}}e^{i\phi_1\sigma_1/2}e^{-i\theta_1\sigma_2/2}|\uparrow\rangle\\[5pt]\nonumber
&=&\;j_{ab}\Big(iV_2^I\,-\,V_1^I\Big),
\end{eqnarray}
\end{widetext}

\noindent where in the end we have taken all angles $\phi_i=\theta_i=0$. Also,  $V_J^I$ is the $I$-th component of the image of the $J$-th unit vector under the rotation $G:=(g_bg_{n_{ba}})^{-1}$, i.e.
\begin{eqnarray}
G\sigma_J G^{-1}\;=\;V_J^I\sigma_I.
\end{eqnarray}

\noindent Furthermore, we have
\begin{eqnarray}\label{Eq:MatrixElement_04}
\frac{\partial^2S}{\partial x_b^I\partial \phi_1}\;&=&\;-\frac{\partial^2S}{\partial x_a^I\partial \phi_1}
\end{eqnarray}

\noindent and
\begin{eqnarray}\label{Eq:MatrixElement_05}
\frac{\partial^2S}{\partial x_b^I\partial \theta_1}\;&=&\;j_{ab}\Big(i V_1^I+V_2^I\Big)\\[5pt]\nonumber
\;&=&\;
-\frac{\partial^2S}{\partial x_a^I\partial \theta_1}\;=\;\frac{1}{i}\frac{\partial^2S}{\partial x_b^I\partial \phi_1}
\end{eqnarray}

\noindent Also, there are, again, equivalent relations for the $\xi_1$ and $\chi_1$ angles, where $a\to a'$, $b\to b'$. Finally, it is not hard to see that the matrix of second derivatives of $x_a^I$
\begin{eqnarray}\label{Eq:MatrixElement_06}
\tilde{H}^{cd}_{IJ}\;:=\;\frac{\partial^2S}{\partial x_c^I\partial x_d^J}
\end{eqnarray}

\noindent at the critical and stationary point coincides precisely with the matrix in the undeformed case - even if $(cd) = (ab)$ or $(a'b')$. The determinants of the Hessian matrix $H$ of the whole integral evaluates to
\begin{eqnarray}
\det(H)\;=\;(j_{ab}j_{a'b'})^{2(n-1)}\det(\tilde H).
\end{eqnarray}

\noindent This is shown in appendix \ref{App:MatricesAndDeterminants}.

From the analysis, it is clear that that the case of more than one crossing is treated in complete analogy, since each link is allowed to partake in at most one crossing. Therefore, the Hessian matrix for the case of more than one crossing can simply be computed by an induction over the number of crossings $C$, and reduced to
\begin{eqnarray}
\det(H)\;=\;\det(\tilde H)\prod_C (j_{ab}j_{a'b'})^{2(n-1)}.
\end{eqnarray}

\subsection{Putting everything together}

We now replace $j_{cd}\to \lambda j_{cd}$, and consider the asymptotic expression for $\lambda\to\infty$. Using the normalized measure on $S^2$ in $\phi,\theta$-coordinates (\ref{Eq:ParameterizationOfSphere}), we get
\begin{eqnarray}
dn_i\;=\;\frac{1}{4\pi}d\phi_id\theta_i\,\cos\theta_i,\; dn'_i\;=\;\frac{1}{4\pi}d\xi_id\chi_i\,\cos\chi_i
\end{eqnarray}

Denote by $\mathcal{B}$ the large-$j$-expression for the undeformed $+$-amplitude (\ref{Eq:VertexAmplitudeFactorised}), and by $\mathcal{B}^\omega$ its deformation. Then, because the critical and stationary points are in one-to-one correspondence, and the Hessian matrix $\det(\tilde{H})$ for the undeformed case can be pulled out of the sum, we have $\mathcal{B}^\omega\;=\;\mathcal{B}\,\mathcal{C}$ with
\begin{eqnarray}\nonumber
\mathcal{C}\;&=&\;\sum_{n=0}^\infty\frac{1}{n!}\left(\frac{4i\omega\sigma(C)}{(1+\gamma)^2}\right)^n
\left(\frac{1}{4\pi}\right)^{2(n-1)}\left(\frac{2\pi}{\lambda}\right)^{2(n-1)}\\[5pt]\nonumber
&&\times
\sum_{I_1,I_2,\ldots, I_n=1}^34^{n-1}\frac{(\lambda j_{ab})^{2n-1}(\lambda j_{a'b'})^{2n-1}}{\sqrt{(j_{ab}j_{a'b'})^{2(n-1)}}}\\[5pt]\nonumber
&&\quad\times
\prod_{i=1}^{n}(\tilde{n}_{ba})^{I_i}\;(\tilde{n}_{b'a'})^{I_i}\\[5pt]\nonumber
&=&\;\sum_{n=0}^\infty\frac{\lambda^{2n}}{n!}(j_{ab}j_{a'b'})^{n}\left(\frac{4i\omega}{(1+\gamma)^2}\right)^n\left(\sum_{I=1}^3(\tilde{n}_{ba})^I\;(\tilde{n}_{b'a'})^I\right)^n\\[5pt]\label{Eq:FinalFormulaPlus}
&=&\;e^{i \omega \lambda^{2}\sigma(C)\vec X_{ab}\cdot \vec Y_{a'b'}}
\end{eqnarray}

\noindent with the vectors $\tilde{n}_{ab}=g_a n_{ab}$, and 
\begin{eqnarray}
\vec X_{ab}\;=\;k_{ab}\,\tilde{n}_{ab},\; \vec Y_{a'b'}\;=\; k_{a'b'}\,\tilde{n}_{a'b'}
\end{eqnarray}

\noindent with $\frac{1+\gamma}{2}k_{cd}=j^+_{ab}=j_{ab}$. This stays finite if, additionally to scaling $j_{cd}$ up by $\lambda$, scaling the deformation parameter as $\omega\to\omega \lambda^{-2}$ at the same time.

This is the computation of the $+$-part, i.e.~$\mathcal{C}^+$. It is noteworthy that $\mathcal{C}^-$ is the same expression, just with a minus sign in the exponential, i.e.~$\mathcal{C}^-= (\mathcal{C}^+)^{-1}$. 

Expression (\ref{Eq:FinalFormulaPlus}) is for one crossing. The case of many crossings is straightforward, however, since we demanded that each edge is part of at most one crossing. For many crossings, one gets
\begin{eqnarray}
\mathcal{C}\;=\;e^{i \omega\sum_C\sigma(C)\vec X_{ab}\cdot \vec Y_{a'b'}}
\end{eqnarray}

\subsection{Relation to the cosmological constant}

We now relate our final result (\ref{Eq:FinalFormulaPlus}) to the cosmological constant. For this, we assume an amplitude in which there are two distinct solutions to the stationary phase equations (\ref{Eq:CritStat01}).\footnote{This seems to be the case whenever the boundary data allows for a unique, non-degenerate $4$-geometry \cite{Bahr:2015gxa, Dona:2017dvf, Bahr:2017eyi}.} We denote these as $g_a^{(i)}$, with $i=1,2$. 

\noindent We denote the asymptotic expression for the undeformed amplitude by
\begin{eqnarray}
\mathcal{A}_\Gamma^\pm\;\longrightarrow\;\mathcal{B}^{\pm}_{(1)}\,+\,\mathcal{B}^{\pm}_{(2)},
\end{eqnarray}

\noindent and from this and (\ref{Eq:FinalFormulaPlus}) one gets that
\begin{eqnarray}\nonumber
\mathcal{A}_\Gamma^\omega\;&\longrightarrow\;\left(\mathcal{B}^+_{(1)}\mathcal{C}_{(1)}^++\mathcal{B}^+_{(2)}\mathcal{C}_{(2)}^+\right)\left(\mathcal{B}^-_{(1)}\mathcal{C}_{(1)}^-+\mathcal{B}^-_{(2)}\mathcal{C}_{(2)}^-\right)\\[5pt]\label{Eq:FullAmplitude}
&\;=\;\mathcal{B}^+_{(1)}\mathcal{B}^-_{(1)}\;+\;\mathcal{B}^+_{(2)}\mathcal{B}^-_{(2)}\\[5pt]\nonumber
&\;\;\; +\;\mathcal{B}^+_{(1)}\mathcal{B}^-_{(2)}\;\mathcal{C}^+_{(1)}\mathcal{C}^-_{(2)}
\;+\;\mathcal{B}^+_{(2)}\mathcal{B}^-_{(1)}\;\mathcal{C}^+_{(2)}\mathcal{C}^-_{(1)}.
\end{eqnarray}

\noindent The terms $\mathcal{B}^+_{(1)}\mathcal{B}^-_{(1)}$ and $\mathcal{B}^+_{(2)}\mathcal{B}^-_{(2)}$ evaluated on the same solution, have been called ``weird terms'', and one can see that they remain unchanged under the deformation of the model. The mixed terms however do get changed, and one has
\begin{eqnarray}\nonumber
\mathcal{C}^+_{(1)}\mathcal{C}^-_{(2)}\;&=&\;\left(\mathcal{C}^+_{(2)}\mathcal{C}^-_{(1)}\right)^{-1}\\[5pt]\nonumber
&=&\;\exp\left(  i\omega\sum_C \sigma(C)\left(  \vec X_{ab}^{(1)}\cdot \vec Y_{a'b'}^{(1)} -  
\vec X_{ab}^{(2)}\cdot \vec Y_{a'b'}^{(2)}\right)  \right)\\[5pt]\label{Eq:VolumeTerm}
&=&\; \exp\left(12i\omega \sum_C \sigma(C)\;*\left(B_{ab}\wedge B_{a'b'}\right)\right).
\end{eqnarray}

\noindent Here $*$ denotes the Hodge dual, $B_{ab}=(\vec{X}_{ab}^{(1)}, \vec{X}_{ab}^{(2)})$ and $B_{a'b'}=(\vec{Y}_{ab}^{(1)}, \vec{Y}_{ab}^{(2)})$ are the bivectors in $\mathbb{R}^4\wedge \mathbb{R}^4\simeq \mathfrak{so}(4)\simeq \mathbb{R}^3\oplus\mathbb{R}^3$ associated to the edges $(ab)$ and $(a'b')$, which are constructed from the two distinct solutions $g_a^{(i)}$. See appendix \ref{App:SelfDualConventions} for details.

\begin{figure}
\includegraphics[scale=0.5]{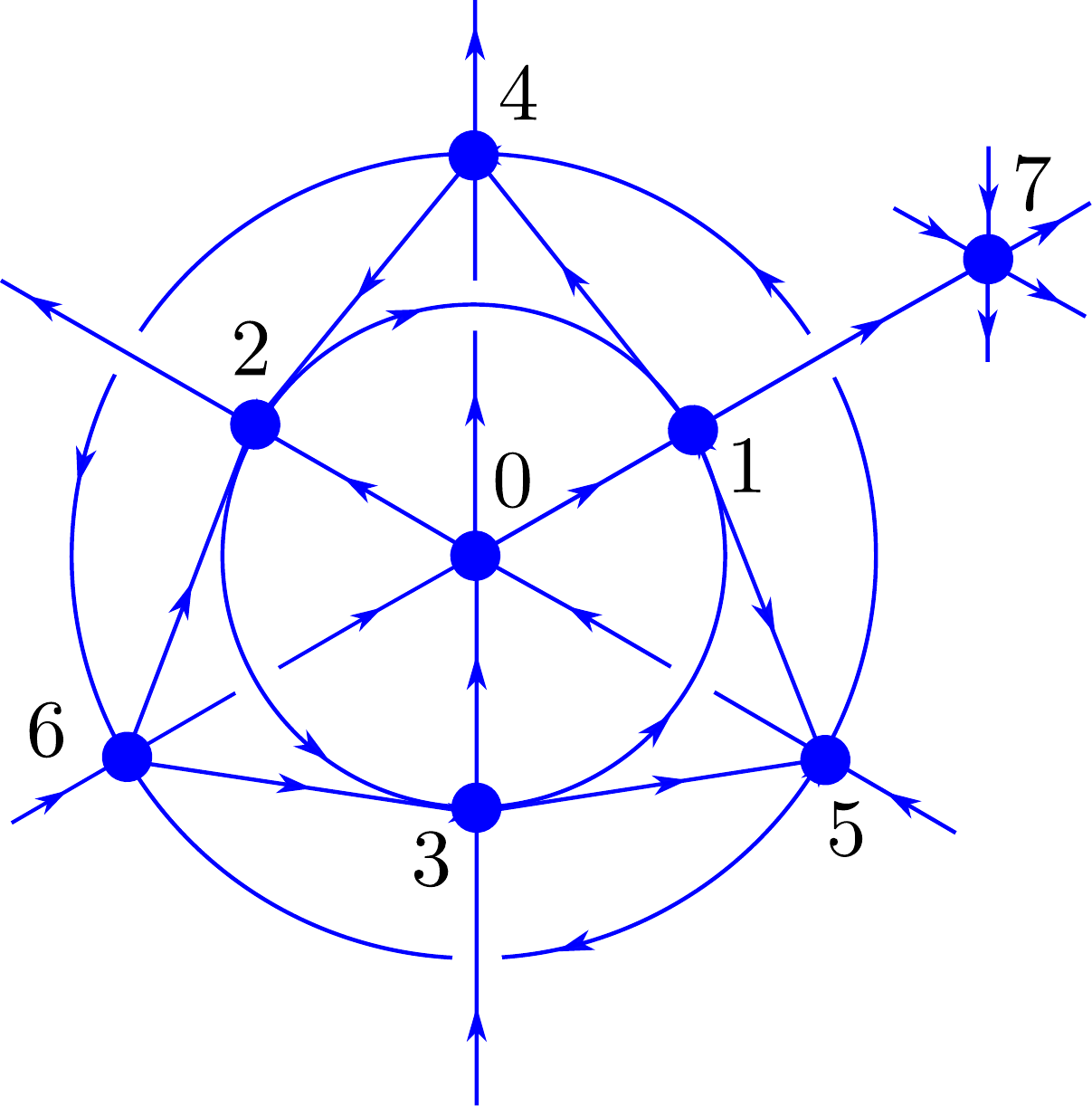}
\caption{The boundary graph of a hypercuboid (or a hyperfrustum). This graph has eight nodes, 24 links, and six crossings. The lines going to infinity all meet at node $7$.}\label{Fig:HypercubicGraph}
\end{figure}

In the case of a $4$-simplex, the expression in (\ref{Eq:VolumeTerm}) has been shown to be proportional to the $4$-volume of such a simplex, given by the boundary data \cite{Han:2011aa}. For the case of the hyperfrustum, which has a boundary graph depicted in figure \ref{Fig:HypercubicGraph}, the critical and stationary equations have been solved in \cite{Bahr:2017eyi}, and the solution can be shown, with the notation from that article, to be
\begin{eqnarray*}
&&\sum_C \sigma(C)\;*\left(B_{ab}\wedge B_{a'b'}\right)\\[5pt]
&&\qquad\;=\;\frac{k(j_1+j_2)}{2}\sqrt{1-\frac{(j_1-j_2)^2}{8k^2}}\;=\;V_{\text{frustum}},
\end{eqnarray*}

\noindent where $V_{\text{frustum}}$ is the $4$-volume of the hyperfrustum. In the case of a hypercuboid, a similar calculation can be carried out. With the notation from \cite{Bahr:2017ajs} and the conventions in appendix \ref{App:SelfDualConventions}, one finds
\begin{eqnarray}
\sum_C \sigma(C)\;*\left(B_{ab}\wedge B_{a'b'}\right)\;=\;\frac{j_1j_6+j_2j_5+j_3j_4}{3},
\end{eqnarray}

\noindent which coincides with $V_{\text{hypercuboid}}$ if the geometricity conditions $j_1j_6=j_2j_5=j_3j_4$ are satisfied. See \cite{Bahr:2017ajs} for a closer discussion of this point, and the relation to the volume simplicity constraints within the EPRL model. 

One can show, indeed, that for convex $4$-dimensional polyhedra $P$ one has in general that
\begin{eqnarray}
V_P\;=\;\sum_C\sigma(C)\; *\left(B_{ab}\wedge B_{a'b'}\right).
\end{eqnarray} 

\noindent A proof will be presented in another publication.

\section{Summary and conclusion}

In this article, we have discussed a generalisation of Han's deformation of the Riemannian signature EPRL model for Barbero-Immirzi-parameter $\gamma\in(0,1)$, to arbitrary vertices. It amounted to introducing an operator depending on a deformation parameter $\omega$, and we have considered the definition for arbitrary graphs as well as the corresponding asymptotic expression of the deformed amplitude $\mathcal{A}_\Gamma^\omega$. This deformation works by introducing an operator for each crossing $C$ of the graph $\Gamma$ in the formula for the amplitude. 

The main statement is that the deformed amplitude $\mathcal{A}_\Gamma^\omega$ has a close connection to $\mathcal{A}_\Gamma$, the undeformed one. Firstly, the equations for the stationary critical points in the asymptotic analysis are in one-to-one correspondence. Also, we could show that the Hessian determinant can be treated, and is just a multiple of the undeformed one. This led to an expression of the asymptotic expression in terms of the original Regge action. In particular, the original expression consists of the so-called weird terms, as well as the cosine of the Regge action. Our analysis shows that the weird terms remain unchanged, while the Regge action is replaced by a term $\Lambda V$, where $\Lambda=-12\omega$, and $V$ is an expression which, if the boundary data is that of a convex, non-degenerate polyhedron, is equal to its volume.
\begin{eqnarray}
\mathcal{A}_\Gamma^\omega\to W+W^*+ \frac{2}{|D|}\cos\left(S_{\text{Regge}} - \Lambda V\right).
\end{eqnarray}

\noindent This way, the deformation provides, in a straightforward way, a generalization of the EPRL-KKL model to include a non-zero cosmological constant $\Lambda$. 

There are two points of note in this analysis:
\begin{enumerate}
\item There are cases in which the boundary data does not describe a vector geometry (in that there are two critical and stationary points), while not describing a $4$d polyhedron. These ``non-geometric'' configurations have been discussed in \cite{Bahr:2015gxa, Dona:2017dvf}, and their presence can be attributed to the insufficient implementation of the volume simplicity constraint. The expression $V$, however, still exists and is non-zero. It is unclear what its geometric interpretation is in that case. 
\item The original EPRL-KKL amplitude $\mathcal{A}_\Gamma$ is defined on a graph $\Gamma$, but does not depend on its knotting class. As a consequence, the physical inner product therefore also does not \cite{Bahr:2010my}. Interestingly, the deformation $\mathcal{A}_\Gamma^\omega$, however, does depend on the knotting of $\Gamma$. This is a property it shares with the quantum group deformations of the model. One can conjecture that this would lead to a physical Hilbert space in which graphs with different knotting classes are not equivalent. This could have interesting physical ramifications. \cite{Rovelli:1987df}
\end{enumerate}

It should also be noted that, while the expression $V$ (\ref{Eq:VolumeTerm}) is a knotting invariant in the asymptotic limit, i.e.~does not depend on the way in which the graph $\Gamma$ is presented on the plane \cite{Bahr:2017ajs}, it is unknown whether the same is true for the quantum amplitude. We will return to this point in a future article.

\section*{Acknowledgements}

This project was funded by the project BA 4966/1-1 of the German Research Foundation (DFG). The authors thank Sebastian Steinhaus for thorough discussions.

\appendix
\section{Bivector conventions}\label{App:SelfDualConventions}

A bivector $B_{ab}=-B_{ba}\in\bigwedge^2\mathbb{R}^4$, $a,b=0,1,2,3$, can be dualized via the Hodge operator
\begin{eqnarray}
(*B)_{ab}\;:=\;\frac{1}{2}\epsilon_{abcd}B_{cd},
\end{eqnarray}

\noindent where indices are raised and lowered with the Kronecker delta $\delta_{ab}$. The Killing form on $\bigwedge^2\mathbb{R}^4$ is taken to be positive definite as
\begin{eqnarray}
\langle B_1,B_2\rangle\;:=\;-\frac{1}{4}\text{tr}(B_1B_2).
\end{eqnarray}

\noindent The isomorphism $\bigwedge^2\mathbb{R}^4\simeq \mathbb{R}^3\oplus\mathbb{R}^3$ 
\begin{eqnarray}
B\;\leftrightarrow\;(\vec{b}^+,\,\vec{b}^-)
\end{eqnarray}

\noindent is given by 
\begin{eqnarray}
b^{\pm,I}\;=\;\frac{1}{2}\left(B_{0I}\pm\frac{1}{2}\epsilon_{IJK}B_{JK}\right)
\end{eqnarray}

\noindent with $I=1,2,3$. The wedge product of two bivectors $B$ and $C$ is defined to be
\begin{eqnarray}
(B\wedge C)_{abcd}\;=\;\frac{1}{24}\epsilon_{abcd}\epsilon^{efgh}B_{ef}C_{gh}.
\end{eqnarray}

\noindent Acting with the Hodge dual on this yields a number which is
\begin{eqnarray*}\
*(B\wedge C)\;&=&\;\frac{1}{24}\epsilon^{efgh}B_{ef}C_{gh}\\[5pt]
&=&\;\frac{1}{12}\left(\vec{b}^+\cdot \vec{c}^+-\vec{b}^-\cdot \vec{c}^-\right),
\end{eqnarray*}

\noindent and which can be regarded as the expression for the 4d volume in the volume simplicity constraint \cite{Engle:2007wy, Bahr:2017ajs}.

\section{Determinant of the Hessian}\label{App:MatricesAndDeterminants}

The Hessian matrix of a term at level $n$ of the expansion (\ref{Eq:ExpansionOfExponential}) is rather involved, and needs to be treated with care. Its matrix elements are given by (\ref{Eq:MatrixElement_01}), (\ref{Eq:MatrixElement_02}), (\ref{Eq:MatrixElement_03}), (\ref{Eq:MatrixElement_04}), (\ref{Eq:MatrixElement_05}), and (\ref{Eq:MatrixElement_06}). In all that follows, remember that the indices $a, b, a', b'$ are not free, but $(ab)$ and $(a'b')$ label the links in the graph which are crossing. If we need free indices from the beginning of the alphabet to indicate nodes, we'll begin with $c, d, \ldots$.

With this, we get that the final Hessian matrix is of the form
\begin{eqnarray}
H\;=\;\left(
\begin{array}{cc}
A&B\\B^T&C
\end{array}
\right),
\end{eqnarray}

\noindent where $C$ is the same Hessian matrix as in the undeformed case, $B$ is the matrix of mixed $X_c^I$ and angle variables, and $A$ is the quadratic matrix of two angle derivatives. We have
\begin{eqnarray}\label{Eq:DeterminantFormula}
\det(H)\;=\;\det(A)\det(C-B^TA^{-1}B).
\end{eqnarray}

\noindent First, we consider the matrix $4(n-1)\times 4(n-1)$-dimensional matrix $A$. It is of the form
\begin{eqnarray}\label{Eq:MatrixA}
A\;=\;\left(
\begin{array}{cc}
D\;&\;0\\0\; & \;D'
\end{array}
\right),
\end{eqnarray}

\noindent where the order of the indices is as:

\begin{eqnarray}\nonumber
\phi_1,\ldots, \phi_{n-1},\theta_1,\ldots, \theta_{n-1},\xi_1,\ldots,\xi_{n-1},\chi_1,\ldots, \chi_{n-1}.\\[5pt]\label{Eq:OrderOfIndices}
\end{eqnarray}

\noindent $D$ is therefore the $2(n-1)\times 2(n-1)$-dimensional matrix which has the form
\begin{eqnarray*}
D\;=\;\frac{j_{ab}}{2}\left(
\begin{array}{cc}
E\;&\;F\\-F\; & \;E
\end{array}
\right),\;
D'\;=\;\frac{j_{a'b'}}{2}\left(
\begin{array}{cc}
E\;&\;F\\-F\; & \;E
\end{array}
\right)
\end{eqnarray*}

\noindent with $E$ and $F$ being $(n-1)\times (n-1)$-dimensional matrices, with
\begin{eqnarray*}
E_{rr}\;&=&\;-2,\qquad r=1,\ldots, n-1,\\[5pt]
E_{r, r+1} \;&=&\; E_{r+1, r}\;=\;1,\qquad r=1, \ldots, n-2,\\[5pt]
F_{r, r+1}\;&=&\;i\qquad r=1, \ldots, n-2,\\[5pt]
F_{r+1,r}\;&=&\;-i, \qquad r=1, \ldots, n-2,
\end{eqnarray*}

\noindent and all other entries being equal to zero. One readily computes
\begin{eqnarray}
\det(D)\;=\;j_{ab}^{2(n-1)},\qquad \det(D')\;=\;j_{a'b'}^{2(n-1)},
\end{eqnarray}

\noindent as well as
\begin{eqnarray}\label{Eq:MatrixD}
D^{-1}\;&=&\;\frac{1}{2j_{ab}}\left(
\begin{array}{cc}
K\;&\;L\\-L\; & \;K
\end{array}
\right), \;\\[5pt]\nonumber
(D')^{-1}\;&=&\;\frac{1}{2j_{a'b'}}\left(
\begin{array}{cc}
K\;&\;L\\-L\; & \;K
\end{array}
\right),
\end{eqnarray}

\noindent with
\begin{eqnarray}\label{Eq:MatrixKL}
K_{rs}\;=\;-\delta_{rs}-1,\qquad L_{rs}\;=\;\left\{
\begin{array}{rl}
i&r<s\\
0&r=s\\
-i&r>s
\end{array}
\right.
\end{eqnarray}

\noindent with $r,s=1,\ldots, n-1$. With this, we get
\begin{eqnarray}\label{Eq:DeterminantMatrixA}
\det(A)\;=\;(j_{ab}j_{a'b'})^{2(n-1)}.
\end{eqnarray}

\noindent Next we are turning our attention to the part $B^TA^{-1}B$. We note that the matrix $B$ is of dimension $ 4(n-1)\times 3N$, where $3N$ is the number of different values of the multi-index $(cI)$, i.e.~$N$ is the number of nodes in the graph. Out of these $3N$ columns, only twelve contain (potentially) nonzero entries, namely $aI, bI, a'I$, and $b'I$, with $I=1,2,3$. These columns are ($u$ runs from $1$ to $4(n-1)$, in the order (\ref{Eq:OrderOfIndices}) given above):
\begin{eqnarray*}
B_{u, (aI)}\;&=&\;x^I\;\delta_{u,1}\;-\;ix^I\;\delta_{u,n}\\[5pt]
B_{u, (bI)}\;&=&\;-x^I\;\delta_{u, 1}\;+\;ix^I\;\delta_{u,n}\\[5pt]
B_{u, (a'I)}\;&=&\;y^I\;\delta_{u,2n-1}\;-\;iy^I\;\delta_{u,3n-2}\\[5pt]
B_{u, (b'I)}\;&=&\;-y^I\;\delta_{u,2n-1}\;+\;iy^I\;\delta_{u,3n-2},
\end{eqnarray*}

\noindent with $x^I=j_{ab}\Big(iV_2^I\,-\,V_1^I\Big)$ and $y^I=j_{a'b'}\Big(i(V')_2^I\,-\,(V')_1^I\Big)$. Denote
\begin{eqnarray}
M\;:=\;B^T A^{-1} B,
\end{eqnarray}
\noindent then it is clear that $M$ is a $3N\times 3N$ matrix, which has zero entries until both row and column index are equal to one of the tweleve combinations $(aI),\ldots (b'I)$ above. Now it is straightforward to show that also
\begin{eqnarray*}
M_{(aI)(a'J)}\;=\;M_{(aI)(b'J)}\;=\;M_{(bI)(a'J)}\;=\;M_{(bI)(b'J)}\;=\;0,
\end{eqnarray*}

\noindent and similarly for other mixed combinations. This is clearly the case, since $A^{-1}$ is block-diagonal, as can be seen from (\ref{Eq:MatrixA}). The potentially nonzero entries are
\begin{eqnarray*}
M_{(aI)(aJ)}\;&=&\;\sum_{u,v=1}^{2(n-1)}B_{u, (aI)}B_{v, (aJ)}(D^{-1})_{uv}\\[5pt]
&=&\;\frac{1}{2j_{ab}}\Big(x^Ix^JK_{11}\;-\;2ix^I x^JL_{11}\;+\;(ix^I)(ix^J)K_{11}\Big)\\[5pt]
\;&=&\;0,
\end{eqnarray*}

\noindent as can be seen from (\ref{Eq:MatrixD}) and (\ref{Eq:MatrixKL}). We also get
\begin{eqnarray*}
M_{(aI)(bJ)}\;&=&\;\sum_{u,v=1}^{2(n-1)}B_{u, (aI)}B_{v, (bJ)}(D^{-1})_{uv}\\[5pt]
&=&\;\frac{1}{2j_{ab}}\Big(-x^Ix^JK_{11}\;+\;2ix^Ix^JF_{11}\;+(-ix^I)(ix^J)K_{11}\Big)\\[5pt]
&=&\;0.
\end{eqnarray*}

\noindent Similar relations hold for $M_{(a'I)(a'J)}$, etc., which lets us conclude that
\begin{eqnarray}
M\;=\;0.
\end{eqnarray}

\noindent With (\ref{Eq:DeterminantFormula}) and (\ref{Eq:DeterminantMatrixA}), this immediately leads to
\begin{eqnarray}
\det(H)\;=\;(j_{ab}j_{a'b'})^{2(n-1)}\det(C),
\end{eqnarray}

\noindent where $C$ is the Hessian matrix of the undeformed case.

\appendix

\bibliography{bibliography}

\begin{thebibliography}{10}

\bibitem{Perez:2012wv}
A.~Perez, ``{The Spin Foam Approach to Quantum Gravity},'' {\em Living Rev.
  Rel.}, vol.~16, p.~3, 2013.

\bibitem{Engle:2007wy}
J.~Engle, E.~Livine, R.~Pereira, and C.~Rovelli, ``{LQG vertex with finite
  Immirzi parameter},'' {\em Nucl. Phys.}, vol.~B799, pp.~136--149, 2008.

\bibitem{Barrett:1997gw}
J.~W. Barrett and L.~Crane, ``{Relativistic spin networks and quantum
  gravity},'' {\em J. Math. Phys.}, vol.~39, pp.~3296--3302, 1998.

\bibitem{Freidel:2007py}
L.~Freidel and K.~Krasnov, ``{A New Spin Foam Model for 4d Gravity},'' {\em
  Class. Quant. Grav.}, vol.~25, p.~125018, 2008.

\bibitem{CubulationSpinFoamThiemann2008}
A.~Baratin, C.~Flori, and T.~Thiemann, ``{The Holst Spin Foam Model via
  Cubulations},'' {\em New J. Phys.}, vol.~14, p.~103054, 2012.

\bibitem{Baratin:2011hp}
A.~Baratin and D.~Oriti, ``{Group field theory and simplicial gravity path
  integrals: A model for Holst-Plebanski gravity},'' {\em Phys. Rev.},
  vol.~D85, p.~044003, 2012.

\bibitem{Bahr:2010bs}
B.~Bahr, F.~Hellmann, W.~Kaminski, M.~Kisielowski, and J.~Lewandowski,
  ``{Operator Spin Foam Models},'' {\em Class. Quant. Grav.}, vol.~28,
  p.~105003, 2011.

\bibitem{Kisielowski:2011vu}
M.~Kisielowski, J.~Lewandowski, and J.~Puchta, ``{Feynman diagrammatic approach
  to spin foams},'' {\em Class. Quant. Grav.}, vol.~29, p.~015009, 2012.

\bibitem{Bahr:2012qj}
B.~Bahr, B.~Dittrich, F.~Hellmann, and W.~Kaminski, ``{Holonomy Spin Foam
  Models: Definition and Coarse Graining},'' {\em Phys. Rev.}, vol.~D87, no.~4,
  p.~044048, 2013.

\bibitem{Kaminski:2009fm}
W.~Kaminski, M.~Kisielowski, and J.~Lewandowski, ``{Spin-Foams for All Loop
  Quantum Gravity},'' {\em Class. Quant. Grav.}, vol.~27, p.~095006, 2010.
\newblock [Erratum: Class. Quant. Grav.29,049502(2012)].

\bibitem{Belov:2017who}
V.~Belov, ``{Poincar\'e-Pleba\'nski formulation of GR and dual simplicity
  constraints},'' 2017.

\bibitem{Bahr:2017ajs}
B.~Bahr and V.~Belov, ``{On the volume simplicity constraint in the EPRL spin
  foam model},'' 2017.

\bibitem{Barrett:2009gg}
J.~W. Barrett, R.~J. Dowdall, W.~J. Fairbairn, H.~Gomes, and F.~Hellmann,
  ``{Asymptotic analysis of the EPRL four-simplex amplitude},'' {\em J. Math.
  Phys.}, vol.~50, p.~112504, 2009.

\bibitem{Fairbairn:2010cp}
W.~J. Fairbairn and C.~Meusburger, ``{Quantum deformation of two
  four-dimensional spin foam models},'' {\em J. Math. Phys.}, vol.~53,
  p.~022501, 2012.

\bibitem{Han:2011aa}
M.~Han, ``{Cosmological Constant in LQG Vertex Amplitude},'' {\em Phys. Rev.},
  vol.~D84, p.~064010, 2011.

\bibitem{Haggard:2015kew}
H.~M. Haggard, M.~Han, W.~Kaminski, and A.~Riello, ``{SL(2,C) Chern-Simons
  Theory, Flat Connections, and Four-dimensional Quantum Geometry},'' 2015.

\bibitem{Dittrich:2018dvs}
B.~Dittrich, ``{Cosmological constant from condensation of defect
  excitations},'' 2018.

\bibitem{Bahr:2016hwc}
B.~Bahr and S.~Steinhaus, ``{Numerical evidence for a phase transition in 4d
  spin foam quantum gravity},'' {\em Phys. Rev. Lett.}, vol.~117, no.~14,
  p.~141302, 2016.

\bibitem{Bahr:2017eyi}
B.~Bahr, S.~Kloser, and G.~Rabuffo, ``{Towards a Cosmological subsector of Spin
  Foam Quantum Gravity},'' {\em Phys. Rev.}, vol.~D96, no.~8, p.~086009, 2017.

\bibitem{Bahr:2017klw}
B.~Bahr and S.~Steinhaus, ``{Hypercuboidal renormalization in spin foam quantum
  gravity},'' {\em Phys. Rev.}, vol.~D95, no.~12, p.~126006, 2017.

\bibitem{Bahr:2015gxa}
B.~Bahr and S.~Steinhaus, ``{Investigation of the Spinfoam Path integral with
  Quantum Cuboid Intertwiners},'' {\em Phys. Rev.}, vol.~D93, no.~10,
  p.~104029, 2016.

\bibitem{Dona:2017dvf}
P.~Donà, M.~Fanizza, G.~Sarno, and S.~Speziale, ``{SU(2) graph invariants,
  Regge actions and polytopes},'' {\em Class. Quant. Grav.}, vol.~35, no.~4,
  p.~045011, 2018.

\bibitem{Bahr:2010my}
B.~Bahr, ``{On knottings in the physical Hilbert space of LQG as given by the
  EPRL model},'' {\em Class. Quant. Grav.}, vol.~28, p.~045002, 2011.

\bibitem{Rovelli:1987df}
C.~Rovelli and L.~Smolin, ``{Knot Theory and Quantum Gravity},'' {\em Phys.
  Rev. Lett.}, vol.~61, p.~1155, 1988.

\end{thebibliography}
\bibliographystyle{ieeetr}

\end{document}